# Liquid Metal Flow Can Be One Clue to Explain the Frequently Observed Fluid-Like Matters on Mars

Jing Liu [1, 2*], Yunxia Gao [2] and Huangde Li [2]

[1] Department of Biomedical Engineering, School of Medicine,
Tsinghua University, Beijing 100084, China

[2] Beijing Key Lab of Cryogenic Biomedical Engineering,
Technical Institute of Physics and Chemistry,
Chinese Academy of Sciences, Beijing 100190, China

*Address for correspondence:
Dr. Jing Liu
Department of Biomedical Engineering,
School of Medicine, Tsinghua University,
Beijing 100084, P. R. China
E-mail address: jliubme@tsinghua.edu.cn
Tel. +86-10-62794896
Fax: +86-10-82543767

**Abstract:** The frequently discovered flooding structure on Mars and other planets has long been an intriguing mystery remained un-disclosed so far. Considering that on Earth, quite a few low melting point liquid metals or their alloy can be candidates of fluid like matters, we proposed here that there might also exists certain liquid metal instead of water or methane alone on Mars or the like planets. Compared with water, such liquid metal would be much easier to stay at the Mars surface because of its low melting point however extremely high evaporation point. Along this theoretical route, quite a few observations on the fluid like matters in former space explorations can be well interpreted. Such hypothesis for the existence of liquid metal on Mars surface does not mean refuting the possibility of water on Mars. This new point would be helpful for planning further exploration of Mars in a sense according to the characters of liquid metal. It at least identifies one more target fluid towards either finding or denying life at outer space. Whether the planet could harbor life in some form or it reaffirms Mars as an important future destination for human exploration still needs serious but not just enthusiasm explorations.

**1. Introduction**

The exploration of Mars has been an important part of the space exploration programs with human fascination. In fact, after Earth, Mars is the planet with the most hospitable climate in the solar system. Dozens of robotic spacecraft, including Orbiters, Landers, and Rovers, have been launched toward Mars since the 1960s to investigate if the conditions there are necessary for life to originate. In recent years, plenty of images from Mars surface have been obtained, especially those for gullies on the red planet transmitted back from the Viking Orbiters. Scientists keep finding that there were many huge gullies on Martian surface (see Fig. 1). For quite a few years, they thought it was the floods of water that carved the deep gullies and traveled thousands of kilometers. The outflow channels and other geologic features seem provided ample evidences that billions of years ago liquid water flowed on the surface of Mars. However, some other scientists did not agree with this claim. Therefore, more explanations remained to be made [1]. For example: the gullies may be attributed to the eroding action from liquid carbon dioxide [2], groundwater [3], brine water [4], water from melting ice [5] or melting snow [6] and so on. Moreover someone even thought that it may be related with wind [7]. Overall, the stories of water on Mars kept to be disputed keenly and so far there are still not many direct evidences to prove the existence of liquid water on Mars.

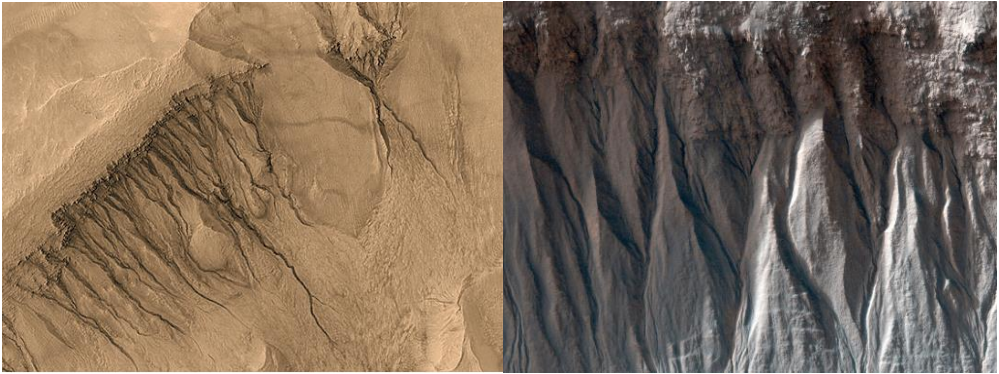

**Fig.1 Huge gullies on the Mars surface (Public Source: NASA/JPL/Malin Space Science Systems)**

Two years ago, when Phoenix landed on the Red Planet, some intriguing images exhibited that certain water like droplets were splashed on the lander's leg (see Fig. 2), which ignites again the claim that liquid water may exist [8]. The issue whether liquid water existed or not on Martian surface keeps calling people's ever increasing attentions. A series of recent news from NASA was frequently disclosed to discuss this issue. Scientists thought such phenomena may be attributed to plenty of perchlorates which were dissolved in water and suppressed its freezing point as low as -70 ℃. However, it was still a guess. This discussion is dedicated to provide certain un-conventional explanations on the observed phenomena.

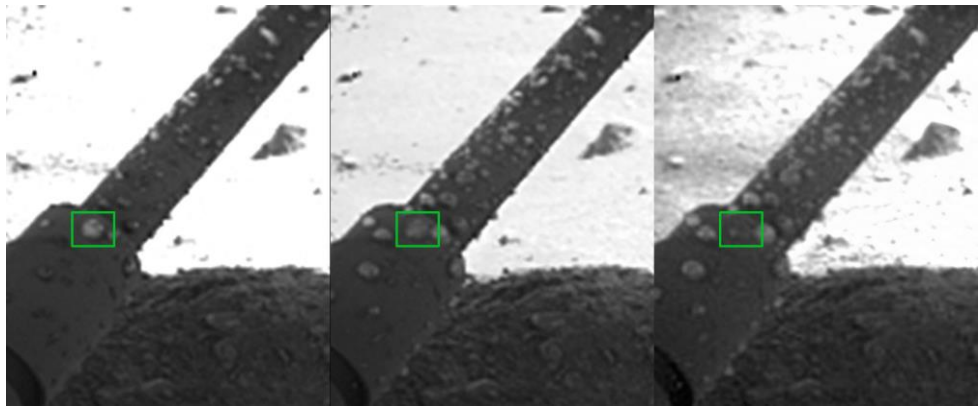

**Fig. 2    Growing water like droplets on Phoenix's legs (Public Source: NASA)**

## 2. Hypothesis of liquid metal on Mars

As is well known, the climate of Mars has obviously cooled dramatically. Although liquid water may exist deeply below the surface of Mars, currently the temperature is just too low and the atmosphere is too thin for the liquid water to remain at the surface.

Here, we would propose that there might exist certain kind of low melting point liquid metal having the same fluidity with water on Mars. Compared with water, the liquid metal would be much easier to stay at the Martian surface because of its low melting point however extremely high evaporation point. In this side, quite a few commonly encountered low melting point liquid metals or their alloy can be such liquid candidates. As shown in table 1, the melting points of Na-K alloys can be easily lower than -10$^o$C, while the melting point for Na-K-Cs alloy already reaches much lower temperature, say -78.2 $^o$C.

**Table 1 Some low melting liquid metal**

| Alloys/Pure metals | $T_m$ ($^o$C) | Alloys/Pure metals | $T_m$ ($^o$C) |
|---|---|---|---|
| $Na_{4.14}K_{22.14}Cs_{73.71}$ | -78.2 | Hg | -39 |
| $Hg_{91.5}Ti_{8.5}$ | -58 | Galinstan | -19 |
| $K_{23}Cs_{87}$ | -48 | $Na_{23.3}K_{76.7}$ | -12.7 |
| $Rb_{13}Cs_{87}$ | -40 | $Na_{22}K_{78}$ | -11 |

**2.1 Possibility of existence of liquid metal**

According to the analysis by NASA scientists on the component of Martian soil gathered from Mars exploration Rovers, Spirit and Opportunity, the soil on Mars was mainly comprised of Na, K, Mg, Ca, Fe, S, Al, Si and other elements [9]. Moreover, there was no more than trace amounts oxygen on Mars. The oxygen existed there mainly as carbon dioxide in the atmosphere and as iron oxides in the soil. Therefore, we proposed that Na and K elements may have a possibility of existence on Mars as a form of alloys. In addition, even if the oxidation reaction occurs, it does not affect the fluidity of liquid metal. The oxidation products would put a coat on liquid metal and prevent the occurrence of further oxidation.

It is well known that the melting points of Na-K alloys are all below 0 $^o$C and their boiling point is generally higher than 700 $^o$C. Compared with water, such liquid metal like Na-K or Na-K-N (Here N stands for one or more other metal elements) will have no problem to exist on Mars surface even though facing the low temperature and thin atmosphere on Mars. Moreover the liquid metal flowing on the sand could not be absorbed in a long period of time despite the extreme dryness of Martian soil.

**2.2 The length of the gullies**

The images snapped by Mars probe showed that the gullies were growing. The length of a 2m wide gully on Martian surface was extended by more than 170m over two and a half Earth years, over the period from Nov 2006 to May 2009, said University of Muenster researchers [10]. Recently, researchers suggested Mars may have seasonal changes and its temperature can be higher than 0 $^{o}$C at warmer season. Therefore it was thought the geological variation of the gullies may be caused by the flow of the mixtures of ice water and sand at Mars' warmer season. However, it is hard to anticipate that the liquid water could exist on Mars for such a long time because of the extreme dryness of soil. Different from this, the liquid metal with larger surface tension may exist steadily. It can thus be deduced that the mystery for the growing of gullies can be well answered by such a natural phenomenon that the liquid metal flows in Martian summer and freezes in Martian winter.

**2.3 The liquid droplets**

Some researchers proposed the water droplets on the Phoenix lander's leg appeared to grow. They seem become bigger and began to flow and mix together finally. Generally speaking, the Mars lander's leg was mainly comprised of Al-Ti alloys. Interesting enough, such phenomenon was frequently observed on liquid metal. According to our previous experiments on the corrosion issues between liquid metal and aluminum alloy [11], it can be found that the corrosion would develop between the aluminum-alloy and liquid gallium (see Fig. 3). The liquid metal droplet spreads around and up-swelled from the flanking corrosion area. Such changing was similar with that of the water like droplets observed on the Phoenix lander's leg. The infiltration action of liquid metal on the aluminum-alloy was rather strong. Only after a few time, the droplet becomes smaller. Following this, we could guess that the change of water like droplets may be originated from the corrosion action between the liquid metal and lander's leg (Al-Ti alloys).

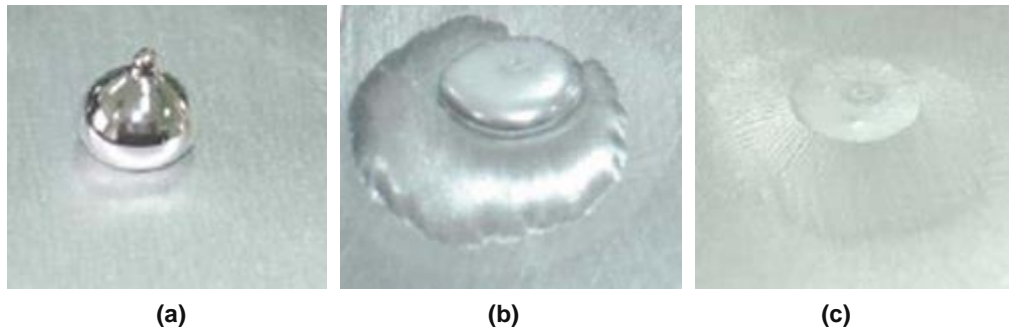

**(a)**              **(b)**              **(c)**

**Fig. 3 Corrosion developments between the 6063 Aluminum-Alloy and liquid gallium: (a) Before corrosion; (b) During corrosion; (c) After corrosion (Source: Experimental Images from the Senior Author)**

## 3. Summary

In a word, the proposal for the existence of liquid metal instead of water alone on Martian surface does not mean refuting the possibility of water on Mars. We hope such point can be helpful for planning further exploration of Mars in a sense according to the characters of liquid metal. It at least identifies one more target towards either finding or denying life at outer space.

Whether the planet could harbor life in some form or it reaffirms Mars as an important future destination for human exploration still needs serious but not just enthusiasm explorations on Mars. "It's a mystery now, but I think it's a solvable mystery with further observations and laboratory experiments," said Alfred McEwen from the University of Arizona[12]. We highly agree with that and firmly believe any doubt about Mars will become clear in the near future.

**References**

1. Yue, Z. Y. *et al*. Earth Science-Journal of China University of Geosciences. **35**, 291-301 (2010).

2. Musselwhite, D. S. *et al*. Geophys. Res. Lett. **28** (7), 1283-1285 (2001).

3. Márquez, A. *et al*. Icarus. **179**(2), 398-414 (2005).

4. Kereszturi, A. *et al*. Icarus. **207**, 149-164 (2010).

5. Costard, F. *et al.* Science. **295**(5552), 110-113 (2002).

6. Christensen, P. R. Nature. **422** (6927), 45-48 (2003).

7. Yue, Z. Y., Xie, H. J. American Geophysical Union, Fall Meeting, abstract No. P31B-0436 (2007).

8. Rennó, N. O. *et al.* J. Geophys. Res. **114**, E00E03 (2009).

9. Gellert, R. *et al*. Science. **305** (5685), 829-832 (2004).

10. http://www.sciam.com.cn/html/tianwen/taikongtansuo/2010/0429/10480.html

11. Deng,Y. G., Liu, J. Appl Phys A **95**, 907-915 (2009).

12. http://www.guardian.co.uk/science/2011/aug/04/strongest-evidence-yet-water-mars#